\documentstyle[12pt]{article}
\def\bib{\bibitem}
\def\be{\begin{equation}}
\def\ee{\end{equation}}
\def\barr{\begin{array}}
\def\earr{\end{array}}
\def\ie{{\it i.e.} }

\def\viz{{\em viz.}}
\def\etal{{\it et al.} }
\def\lsim{\:\raisebox{-0.5ex}{$\stackrel{\textstyle<}{\sim}$}\:}
\def\gsim{\:\raisebox{-0.5ex}{$\stackrel{\textstyle>}{\sim}$}\:}

\def\EE{$E \bar{E}$}
\def\NN{$N \bar{N}$}
\def\gev{\: {\rm GeV} }
\def\ptsl{$\not${\hbox{\kern -2.5pt $p_T$}} }
\def\vli{v^{(L)}_i}
\def\vlj{v^{(L)}_j}

\def\vei{v^{(e)}_i}
\def\vej{v^{(e)}_j}
\def\ali{a^{(L)}_i}
\def\alj{a^{(L)}_j}

\def\aei{a^{(e)}_i}
\def\aej{a^{(e)}_j}
\def\ptsl{p_T \hspace{-2.3ex}/\;}

\def\ib#1,#2,#3{       {\it ibid.\/ }{\bf #1} (19#2) #3}
\def\ap#1,#2,#3{       {\it Ann.~Phys.~(NY)\/ }{\bf #1} (19#2) #3}
\def\ijmp#1,#2,#3{     {\it Int.~J.~Mod.~Phys.\/ } {\bf A#1} (19#2) #3}
\def\mpl#1,#2,#3 {     {\it Mod.~Phys.~Lett.\/ } {\bf A#1} (19#2) #3}
\def\np#1,#2,#3{       {\it Nucl.~Phys.\/ }{\bf B#1} (19#2) #3}
\def\npps#1,#2,#3{     {\it Nucl.~Phys.~B (Proc.~Suppl.)\/ }{\bf B#1}
                             (19#2) #3}
\def\plb#1,#2,#3{      {\it Phys.~Lett.\/ }{\bf B#1} (19#2) #3}
\def\pr#1,#2,#3{       {\it Phys.~Rev.\/ }{\bf D#1} (19#2) #3}
\def\prep#1,#2,#3{     {\it Phys.~Rep.\/ }{\bf #1} (19#2) #3}
\def\prl#1,#2,#3{      {\it Phys.~Rev.~Lett.\/ }{\bf #1} (19#2) #3}
\def\pro#1,#2,#3{      {\it Prog.~Theor.~Phys.\/ }{\bf #1} (19#2) #3}
\def\rmp#1,#2,#3{      {\it Rev.~Mod.~Phys.\/ }{\bf #1} (19#2) #3}
\def\sp#1,#2,#3{       {\it Sov.~Phys.-Usp.\/ }{\bf #1} (19#2) #3}
\def\zpc#1,#2,#3{      {\it Zeit.~f\"ur Physik\/ }{\bf C#1} (19#2) #3}
\begin{document}
\setcounter{page}{0}
\renewcommand{\thefootnote}{\fnsymbol{footnote}}
\thispagestyle{empty}
\vspace*{-1in}
\begin{flushright}
CERN-TH/95--306 \\
MPI-PTh/95-135\\[2ex]
{\large \tt hep-ph/9512317} \\
\end{flushright}
\vskip 45pt
\begin{center}
{\Large \bf Fourth-Generation Leptons at LEP2}

\vspace{11mm}
\bf
Gautam Bhattacharyya\footnote{ gautam@cernvm.cern.ch}

\vspace{10pt}
{\em Theory Division, CERN, CH-1211 Geneva 23, Switzerland.}

\vspace{12pt}
\rm and

\vspace{12pt}
\bf Debajyoti Choudhury\footnote{debchou@mppmu.mpg.de}\\

\vspace{10pt}
{\em Max--Planck--Institut f\"ur Physik, Werner--Heisenberg--Institut,\\
F\"ohringer Ring 6, 80805 M\"unchen,  Germany.}

\vspace{50pt}
{\bf ABSTRACT}
\end{center}

\begin{quotation}

{}From non-observation at LEP1, a lower mass limit of
45 GeV has been established on any additional sequential fermion
beyond the three generations. Precision measurements have further
constrained the number of such additional generations, either
degenerate or nearly so, to be at most one.
LEP2, an energy-upgraded version of LEP1,
would provide greater mass-reach in the  search for such
particles. We study the pair-production  of  fourth-generation
leptons for various LEP2 energy options.
We find that in most cases such particles could be discovered/ruled out
up to the kinematic limit.

\end{quotation}

\vspace{130pt}
\noindent
\begin{flushleft}
CERN-TH/95--306\\
November 1995\\
\end{flushleft}

\vfill
\newpage
\setcounter{footnote}{0}
\renewcommand{\thefootnote}{\arabic{footnote}}

\setcounter{page}{1}
\pagestyle{plain}
\advance \parskip by 10pt
\section{Introduction}
Despite the vindication of the Standard Model (SM) at LEP1, a few questions
remain unanswered, not the least of which concerns the twin issues of
fermion masses and family replication. While, as yet, there exists no
understanding of either issue, there have been some speculations to the
effect that these might be interrelated~\cite{4th family,4th-new,exotic}.
Furthermore, the existence of such particles are likely to have a
significant impact even on low-energy observables, {\em e.g.} particularly
in the context of CP violation in the $B$-system~\cite{4th-cp}.
The issue of exploring the existence of new fermions at LEP2 thus turns
out to be imperative.

Possible new quarks or leptons can be subdivided into two categories :
sequential (\ie with gauge quantum numbers identical to the SM fermions)
 or  exotic~\cite{exotic}. The latter class would include all such
fermions that have no analogue in the SM.

The most model-independent bounds on fermions beyond the SM can be inferred
from their non-observation at LEP1.
Unless their couplings to the $Z$ are highly suppressed,
for all such fermions~\cite{pdg}: $m_F > 45 \gev$.
At the Tevatron, on the other hand, quark pair-production far outstrips
that for leptons. Assuming that a heavy quark decays within the detector
via its neutral current or charged current mixing with the
standard quarks, one obtains a stronger bound
$m_Q > 85~\gev$~\cite{pdg,bm_dp}, which is likely to improve with new data.
As lepton production at the Tevatron proceeds through
Drell--Yan processes or through weak-gauge boson fusion, the
corresponding limits are expected to be weaker. This analysis
has not yet been reported though.
Over and above these direct bounds, the bound  on the oblique parameter
$S$ (equivalently $\epsilon_3$) from the precision electroweak measurement
restricts the number of additional degenerate chiral generations
to just one \cite{LEP-precision}.
It is thus quite clear that any addition to the fermionic sector of the SM
cannot be too arbitrary and hence all search strategies should be devised
keeping the existing bounds in mind. In this article we examine the
possibility of discovering new fermions at the
forthcoming upgrade of the LEP machine at CERN. As it is supposed to
operate at $\sqrt{s} = 140, 176, 192$  and (hopefully) 205 GeV,
 heavy quark pair-production is interesting only for the last two
energy options. Single
production in association with a light quark is  ruled out (at an
observable level) from the severe bounds on
flavour-changing neutral currents (FCNC) involving light quarks.
 Hence, in the present work we concentrate on heavy leptons only.

Now, the search pattern for exotic particles~\cite{djouadi}
necessarily differs from that for sequential fermions.
Although the mass reach is much more for the former, as they may be singly
produced, the production cross section suffers a suppression
from the low-energy bounds on lepton mixing~\cite{bounds:2-fam}.
Pair production of the exotics has been considered
in ref.~\cite{djouadi} and we shall thus confine ourselves to
sequential leptons.

\begin{figure}[htbp]
\vskip 4.2in\relax\noindent\hskip -0.7in\relax{\includegraphics{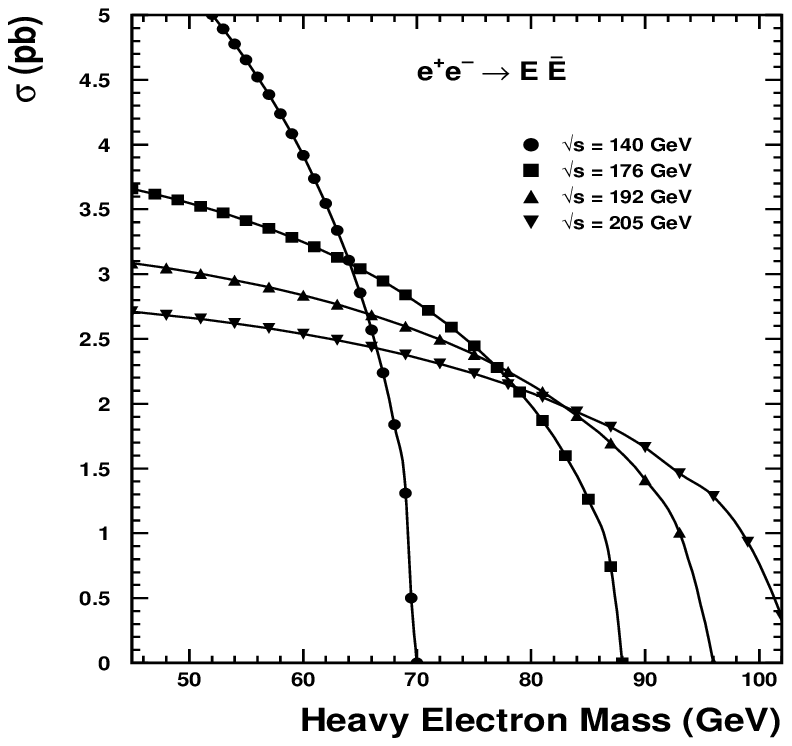}}
          \relax\noindent\hskip 3.5in\relax{\includegraphics{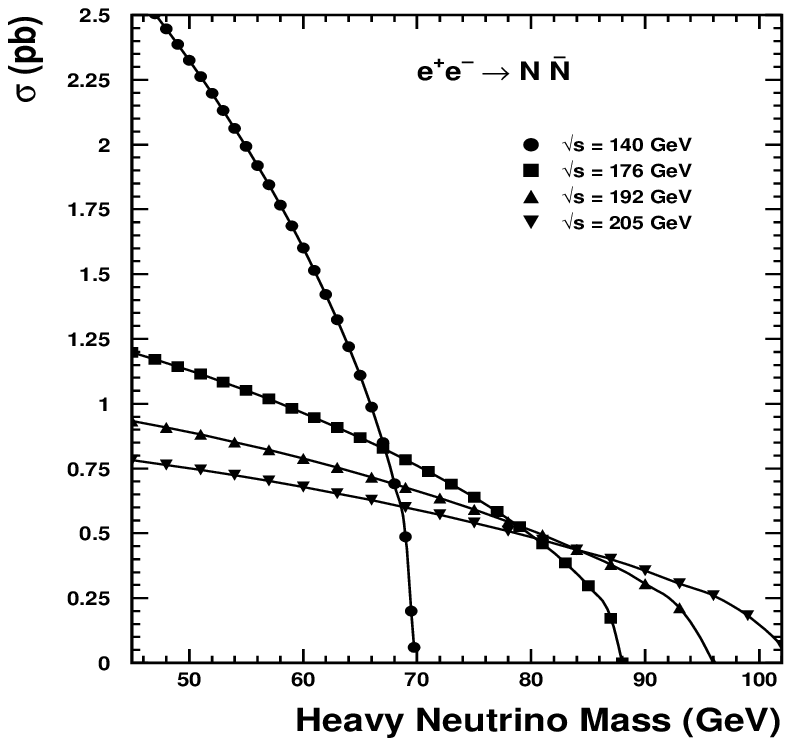}}
\vspace{-18ex}
\caption{ {\em The total cross section (at various c.m. energies)
 for $E^+ E^-$ and $N \bar{N}$ production at LEP2.} }
\label{fig-cs}
\end{figure}
Since we restrict our analysis to sequential fermions, it is clear that
there are no tree-level FCNCs in the theory.
The production of a heavy charged lepton pair (\EE)
at LEP2 then proceeds primarily
through a Drell--Yan-like mechanism (\viz exchange of
$s$-channel $\gamma/Z$).
For \NN\  production, the $\gamma$ channel is obviously absent.
There is an additional diagram, though, involving a  $t$-channel $W$
exchange. However since this contribution to the amplitude involves the
$eN$ mixing, which is already restricted to be very small
$(\sin^2 \theta_{eN} \sim \sin^2 \theta_{\nu N}\lsim 0.005)$,
the $t$-channel exchanges
can be essentially neglected. The
production cross-section,
can then be  calculated in a straightforward
manner and is given in the appendix.
In Fig.~\ref{fig-cs}, we display the total cross section for the
pair-production of both $N$ and $E$ as a function of their
masses\footnote{Here, as in most of the following discussion, we
neglect the effect of initial state radiation (ISR). As $N\bar{N}$
production proceeds mainly through an $s$-channel $Z$-exchange, the effect
of ISR is expected to be severe on account of the tendency to return to
the $Z$-pole.
However, from the quantitative analysis of ref. \protect\cite{shev}, we
see that the effect is actually not severe enough to change our
qualitative predictions. For charged leptons ($E$),
the effect is even smaller.}. That the \EE\ cross section is typically
larger than the corresponding \NN\ cross section is due to the presence
of the extra ($\gamma$--exchange) diagram for the former.
As can be  easily
seen, the cross-section is quite large almost up to the kinematic
limit of LEP2.
Were all the produced leptons available for detection, this would
be a very good theatre for discovery.
It is thus meaningful to consider the potential backgrounds
and evaluating prospect of detecting such fermions at LEP2.

Since an absolutely stable heavy lepton (charged or neutral) is
disfavoured on  astrophysical/cosmological grounds~\cite{cosmo}, any
such new particle must necessarily decay. Although a very small width
can evade this bound, such particles would be effectively stable as far
as the collider experiments are concerned. While a stable $N$ would
go undetected in the LEP experiments, it might just be possible to
detect a stable $E$ through a calorimetric measurement. We shall, however,
assume that these particles, if produced, decay within the detector volume.
Since tree-level FCNC's are absent in the simplest
fourth generation scenario, all decays proceed through the charged
current interaction.
Current bounds on possible lepton mixing suggests the heavy lepton
$L \; (= E,N)$ would, if allowed to kinematically,
tend to decay predominantly into its isospin
partner. Rather than look for such cascade decays, it makes
more sense to concentrate instead on the lighter of $E,N$,
especially since we
shall show that both such particles can be discovered for almost
all of the parameter space. This particle may then
decay only into fermions of the first three generations.
Decay within the detector volume is assured as long as the mixing angle
is $\theta_{lN}\;(\theta_{\nu E}) \gsim 10^{-6}$.

The rest of the article is planned as follows:
in section~\ref{sec:L} we discuss the different decay modes of $L(E,N)$,
the kinematic cuts implemented in our analysis, the SM backgrounds;
we then highlight the post-cut efficiencies. We draw our conclusion in
section 3. In the appendix, we list all the relevant formulae.

\section{The Signal} \label{sec:L}
According to our criterion, the heavy charged lepton $E$
would decay into one of the lighter neutrinos and a $W$ (on- or off-shell),
\ie
\be
    E^- \rightarrow \nu_i W^{- (\ast)} \qquad  {\rm and} \qquad
    E^+ \rightarrow \bar\nu_i W^{+ (\ast)}\ ,
        \label{E-2b}
\ee
where $i = 1,2,3$. The individual mixing
matrix elements are of no consequence as long as at least one of them
$\gsim 10^{-6}$. When considering decays of $E$ into neutrinos of
different flavour, care must be taken though that the choice of
parameters does not lead to unacceptably large leptonic
FCNC's~\cite{bounds:2-fam}. However, since such decay modes do not
afford us any particular experimental advantage, we shall assume
that $E$ decays to only one specific light neutrino flavour.
The $W$'s may then go into either hadronic or leptonic
channels. The final state is thus one of
\begin{enumerate}
\item $\nu_i \bar\nu_i + q_a \bar{q_b} q_c \bar{q_d} $,
\item $\nu_i \bar\nu_i + q_a \bar{q_b} ( l_k \bar\nu_k + \bar{l_k} \nu_k )$,
\item $\nu_i \bar\nu_i + l_k \bar\nu_k \bar{l_n} \nu_n $.
\end{enumerate}
Of the three modes, the last one is of little use. The signal
would be overwhelmed by the background from $e^+ e^- \rightarrow W^+ W^-$
with each $W$ decaying leptonically. The second option is somewhat better.
We shall, however, confine ourselves to the first and
the most promising  channel with the signal
\be
e^+ e^- \rightarrow E^+ E^- \rightarrow {\rm 4 \ jets} + \ptsl 
      \label{E-signal}
\ee
where 
$\ptsl$ represents the missing transverse momentum.

We now turn to the case where the fourth-generation neutrino $N$
is lighter
than its charged counterpart. Unlike the charged particle, $N$ could have
a Majorana mass too. Apart from leading to interesting possibilities
in low-energy neutrino phenomenology~\cite{majorana}, such an eventuality
leads to the beautifully clean signal of like-sign dileptons with
hadronic activity but without any missing momentum. The  backgrounds
are twofold : ($i$) cascade decays of a heavy quark pair (say $b \bar{b} \to
c e^- \bar\nu \bar{c} q_i \bar{q_j} \to e^- e^- \nu \nu + \; {\rm jets}$)
or ($ii$) effects like $B$--$\overline{B}$ mixing. Although both these
backgrounds ostensibly lead to missing momentum, these are still
relevant as the neutrinos may be slow, or the transverse component
($\ptsl$) 
of the total missing momentum may be small. These can however be
easily eliminated by a combination of isolation cuts and imposition
of a upper bound on missing momentum. Having argued for the
ease of detection in the Majorana $N$ case, we
shall desist from discussing it any further
and will rather concentrate on the case where $N$ is a Dirac particle.
The primary decay vertices then are
of the form
\be
    N \rightarrow l_i^- W^{+ (\ast)} \qquad {\rm and} \qquad
    \bar N \rightarrow l_j^+ W^{- (\ast)}\ ,
        \label{N-2b}
\ee
where $i,j = 1,2,3$. Some of
the most striking signals would emanate in the case $i \neq j$ in
eq. (\ref{N-2b}). However,
such a scenario will induce leptonic FCNC's  at the one-loop level.
The non-observation of such effects can be translated to
considerably strong constraints~\cite{bounds:2-fam} in such
scenarios.
We assume henceforth that $N$, like $E$, has charged-current coupling
with only {\it one} of the light flavours. Even with this simplification,
two further possibilities exist:
($i$) $N$ decays mainly into $e$ or $\mu$.
($ii$) $N$ decays mainly into $\tau$.
Since $\tau$'s decay within the detector volume, the signal profile changes
in an essential manner. Consequently, these two cases must be discussed
separately. We shall, for the present, concentrate on the first case as
$\tau$-identification is tricky, especially in the presence of
hadrons.

Quite analogous to the case of $E$,
the $W$'s in eq. (\ref{N-2b}) may  go into either hadronic or leptonic
channels. The final states are of the following types:
\begin{enumerate}
\item $l_i \bar l_i + q_a \bar{q_b} q_c \bar{q_d} $,
\item $l_i \bar l_i + q_a \bar{q_b} ( l_k \bar\nu_k + \bar{l_k} \nu_k ), $
\item $l_i \bar l_i + l_k \bar\nu_k \bar{l_n} \nu_n. $
\end{enumerate}
The first two modes are better suited for discovery (the first especially,
as it affords easy mass reconstruction). Although the last of the three
can also be useful, we shall not consider it here.

Having identified our final states, it now remains to calculate the
effective transition matrix element. We do this within the narrow width
approximation (albeit keeping track of the spin-spin correlations)
and this derivation is presented in the appendix. It turns out that, on
account of the peculiar kinematics of the system and our choice of
observables, the spin correlations are
not particularly significant.

At this stage it must be pointed out that the above-mentioned signals are
not necessarily the only viable ones. In fact, quite a few
studies~\cite{shev,lep2} have focussed on an $N \bar{N}$ production signal
of isolated leptons accompanied by hadronic activity. While the effective
signal strength is higher for such configurations, the backgrounds are
larger too, and consequently special kinematic cuts (for example, removal
of the $ZZ$ or $Z\gamma^\ast$ backgrounds) are necessary. We,
on the other hand, have chosen final states such that the corresponding
SM backgrounds are easily reduced to rather innocuous levels.

\subsection{Kinematic cuts} \label{kinem}
A multitude of cuts are necessary both for experimental reasons and
to suppress all possible SM backgrounds. To list:
\begin{enumerate}
\item Each quark in eq. (\ref{E-signal}) should be sufficiently away
from the beam pipe and should carry sufficient energy and
transverse momentum.  We require
   \be
         E_j> 10 \gev, \qquad p_{Tj} > 5 \gev, \qquad
        10^\circ < \theta_j  < 170^\circ.
             \label{E:cut:pT,ang}
   \ee

\item Since we do only a parton-level simulation, we must ensure that
the angular separation between any two quarks
 must be large enough for them to
lead to recognisably different jets. To be conservative,
we require, for any two jets,
   \be
        \theta_{jj} > 30^\circ.
             \label{E:cut:ang_jj}
   \ee
This cut also eliminates potentially large contributions from collinear
singularities in the SM matrix elements.

\item
  For final states with charged leptons, we require
\be
        p_{Tl} > 5 \gev, \qquad
        10^\circ < \theta_l < 170^\circ.
        \label{lepton-cut}
\ee
The angular cut also eliminates collinear singularities for the case
$l = e$.
On the other hand, for the \EE\ case,
where the final state contains neutrinos, we demand
that the missing transverse momentum\footnote{Note that missing
longitudinal
momentum is not necessarily a good signal, as it can possibly be faked by
ISR.}
should be sufficiently large as to be observable :
   \be
        \ptsl > 10 \gev.
             \label{E:cut:pt_miss}
   \ee

\item
We further impose the following isolation cuts (whenever applies
depending on the final states)
\be
\barr{rcl}
\theta_{jl}, \theta_{ll}, \theta_{j\not{p}} > 20^\circ.
             \label{E:cut:ang_jp}
\earr
   \ee
Apart from facilitating proper detection, this also serves to reduce the
background contribution where the lepton (or neutrino) arises from a
heavy quark decay. In addition, the cut on $\theta_{ll}$ removes
electromagnetic collinear singularities in the SM matrix elements.

\end{enumerate}

\subsection{Backgrounds}
As our final states typically comprise of  six fermions, power counting
(in terms of the coupling constants) would naively dictate that the
expected SM backgrounds be small. A possible  counterargument
could be that the large number of diagrams  contributing
to such processes might enhance the effects.
In fact, it is precisely this (\viz, a large number of graphs)
that makes an accurate
calculation of the SM backgrounds an extremely difficult task. There
does not exist any computational
tool that  calculates the full $2 \to 6$ matrix elements
in an acceptable time-frame. One should also bear in mind
the possibility
that the backgrounds may be enhanced on account of the
presence of resonant processes and/or collinear emissions.

However, very reasonable estimates may be made
by classifying the possible Feynman diagrams in terms of
well-identifiable ``subprocesses'', \viz
$e^+ e^- \rightarrow e^+ e^- W^{(\ast)} W^{(\ast)}$ with the
$W^{(\ast)}$'s decaying
hadronically, or $e^+ e^- \rightarrow \nu \bar\nu q \bar{q}$ followed by
$q \bar{q} \to 4 \;{\rm jets}$. Such piecemeal calculations are expected
to give an estimate not drastically different from
the full $2 \to 6$ calculation\footnote{The caveat is that such an
analysis ignores those diagrams that cannot be broken down in terms of
simple `subprocesses'. But then, these are truly higher-order in the
coupling
constants and the corresponding contributions are small.}.

These reduced processes were calculated with the help of the helicity
amplitude package MadGraph~\cite{madgraph} and some of them
were counterchecked against
an  independent analysis performed with GRACE~\cite{grace}.
With the adoption of the kinematic cuts listed in the next subsection,
each of these individual contributions is reduced to well below 1 fb, and
it may thus be safely assumed that
 the SM backgrounds  to our signals are finally of no consequence.

\subsection{The Efficiencies}
\begin{figure}[htbp]
\vskip 4.2in\relax\noindent\hskip 1.0in\relax{\includegraphics{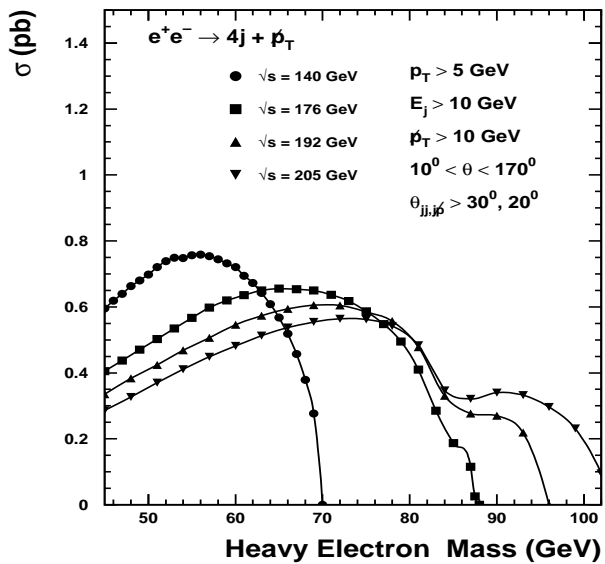}}
\vspace{-18ex}
\caption{ {\em The effective cross section (at various c.m. energies)
for $E^+ E^-$ production after imposition of the cuts of
section~\protect\ref{kinem}. } }
\label{fig-E-p.c}
\end{figure}
\begin{figure}[h]
\vskip 4.2in\relax\noindent\hskip -0.7in\relax{\includegraphics{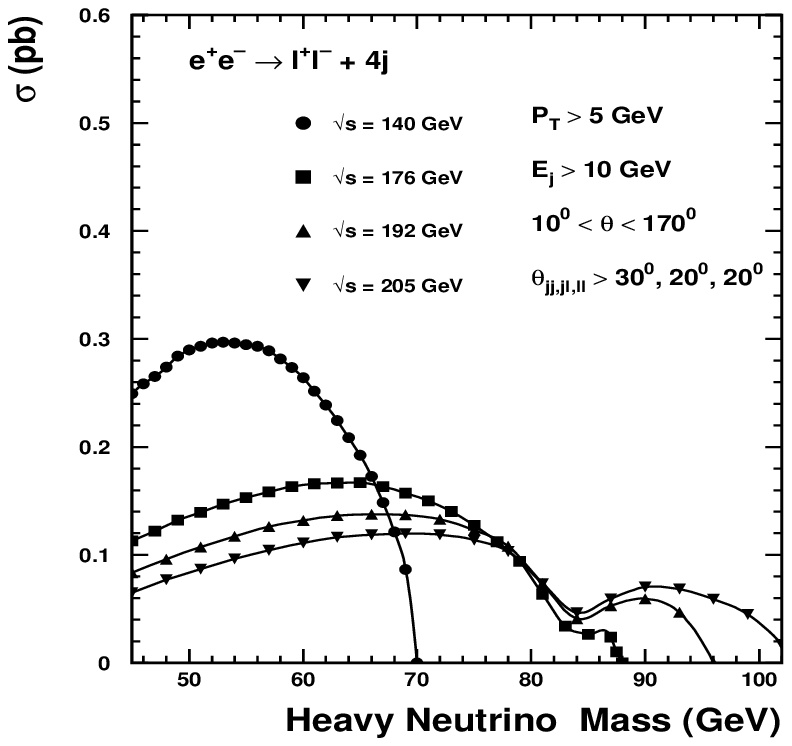}}
          \relax\noindent\hskip 3.5in\relax{\includegraphics{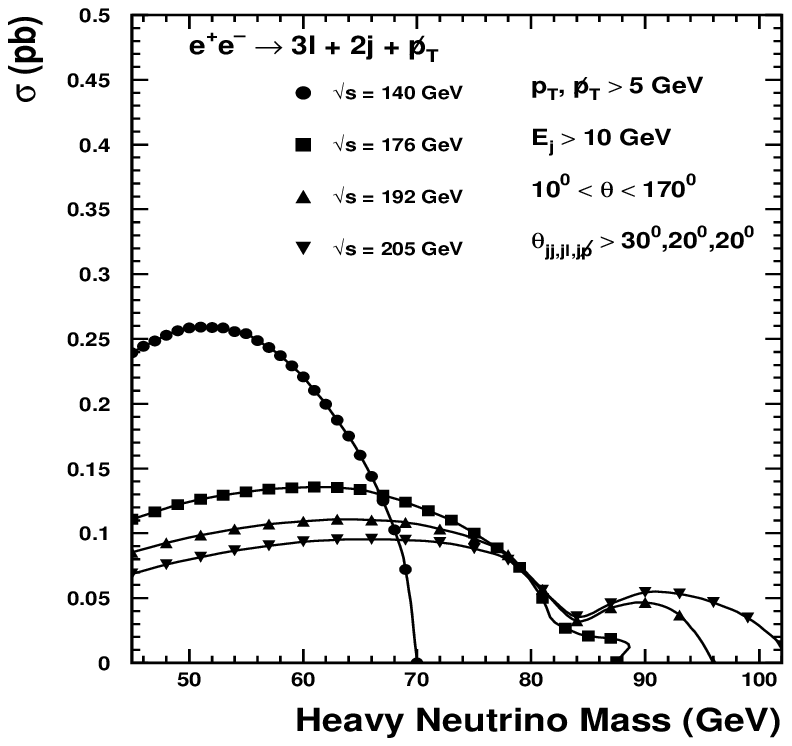}}
\vspace{-18ex}
\caption{ { \em The effective cross section (at various c.m. energies)
for pair-produced $N$'s decaying into two particular channels.
The cuts of section~\protect\ref{kinem} have been imposed. } }
\label{fig-N-p.c}
\end{figure}

While the cuts imposed above serve to eliminate all the SM backgrounds,
they are not particularly severe on the signal. In Fig. \ref{fig-E-p.c},
we show the effective signal strength (\ie after the imposition of the
above-mentioned cuts) for $E^+ E^-$ production with both $W^{(\ast)}$s
decaying hadronically. On the other hand, Fig. \ref{fig-N-p.c} shows
the same for two different decay modes of the $N \bar{N}$ pair. The
post-cut efficiencies
(without folding any detector effects) depend on $m_L$
($L = E, N$) and
can be deduced by comparing these figures with those in Fig. \ref{fig-cs}.
Since the  phase-space volume available to the decay products from
$L(E,N)$ grows as $m_L$, it is obvious that
for low $m_L$, the twin requirements of minimum jet (lepton)
energy and the isolation cuts would substantially reduce the efficiency.
However, this is not a cause for worry as
the production cross-sections are rather large. On the other hand,
for $m_L \gsim m_W$, $L$ prefers a two-body decay (into
an on-shell $W$) rather than a genuine three-body decay. Consequently,
for $m_L$ just above $m_W$ the primary lepton is left with very little
momentum and such configurations are eliminated primarily by the cuts of
eq. (\ref{lepton-cut}). This leads to
the dips in Figs. \ref{fig-E-p.c} and \ref{fig-N-p.c}.
As can easily be ascertained, even after the imposition
of such strong cuts, the effective cross sections are large enough to
warrant a discovery claim right up to the kinematic limit for the
expected luminosity of 300 pb$^{-1}$. On the other hand,
if an integrated luminosity
of 10 pb$^{-1}$ is achieved for the run at $\sqrt{s} = 140 \gev$, it might
be possible to rule out $m_E \lsim 60 \gev$.

\begin{figure}[h]
\vskip 4.2in\relax\noindent\hskip -0.7in\relax{\includegraphics{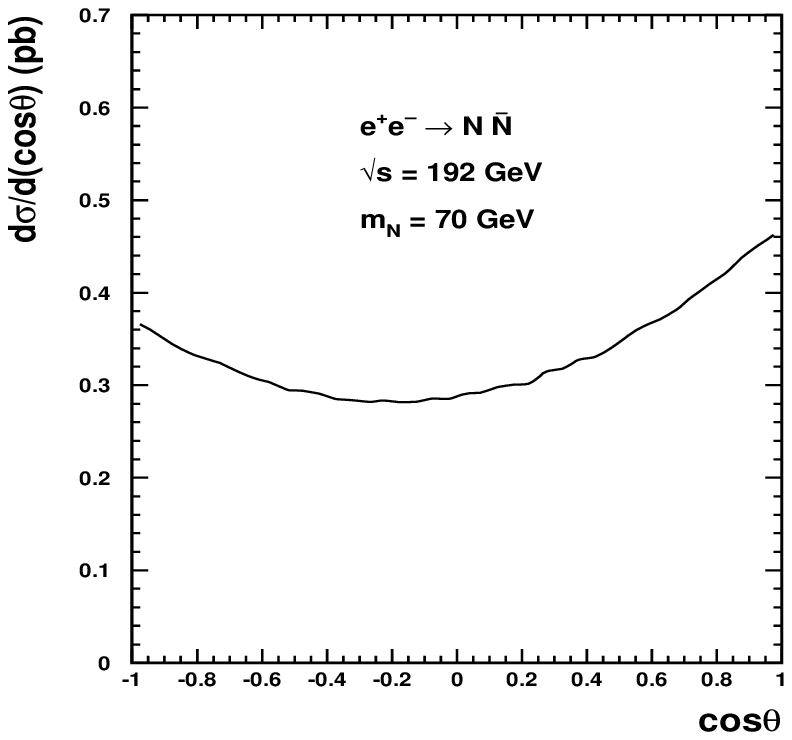}}
          \relax\noindent\hskip 3.5in\relax{\includegraphics{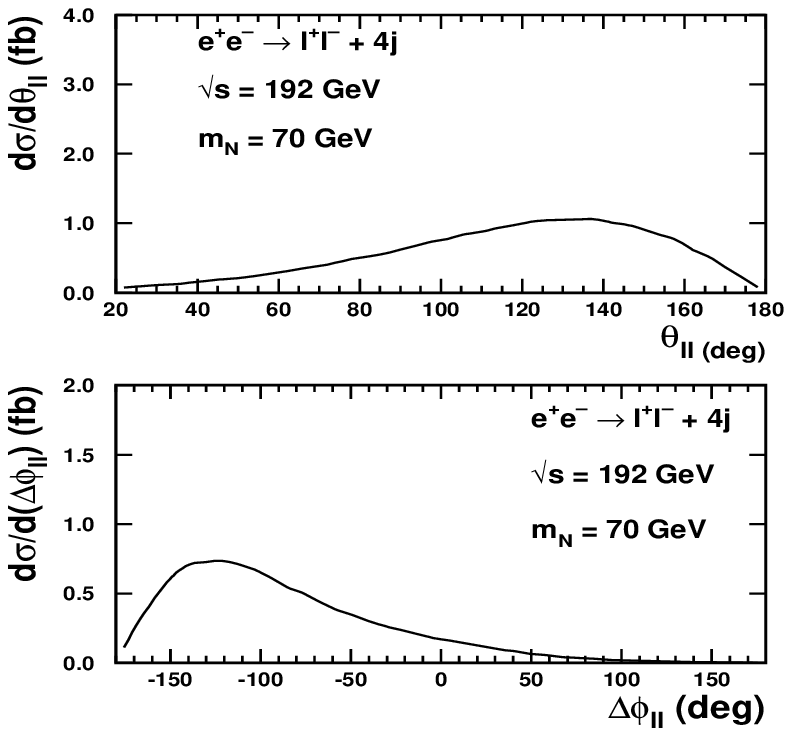}}
\vspace{-18ex}
\caption{ {\em Angular distributions for $N \bar{N}$ production. ($a$) The
variation with the production angle {\em before} the cuts are
imposed. ($b$) The angular distributions ({\em after} imposition of cuts
of section~\protect\ref{kinem}):
(i) the opening angle and (ii) the difference of azimuthal angles)
for the leptons from the primary decay vertices of $N$ and $\bar{N}$. } }
\label{angular}
\end{figure}

Although the cross sections are smaller for $N \bar{N}$, than for
$E^+ E^-$, there are some advantages associated with the first, not the
least of which being mass reconstruction and a window to the angular
distribution (see Fig.~\ref{angular}a).
The latter can differentiate between heavy neutrinos with
different gauge quantum numbers through their polar angle distributions.
Although the imposition of the cuts tends to distort
the distribution, to a certain extent
the characteristic signatures are still preserved.
In Fig. \ref{angular}b, we concentrate on the case of the $(l^+ l^- + 4j)$
signal and show the distribution and function of two relevant
angles : ($i$) the opening angle of the two leptons ($\theta_{ll}$),
and ($ii$) their azimuthal separation ($\Delta\phi_{ll}$).
The ``peaks'' are characteristic, though they
tend to get flattened as $m_N/ \sqrt{s}$ increases.

\section{Conclusions} \label{sec:concl}
We have examined the possibility of observing a fourth-generation
lepton in the forthcoming runs at LEP2. The prognosis for
quarks is not favourable. Only the runs at $\sqrt{s} = 192$ and 205 GeV
may explore mass ranges not ruled out as yet, and that window too
might already be closed from the upcoming Tevatron data. On the other hand,
both charged and neutral leptons can be discovered (or ruled out)
up to the kinematic limit. Although we attempt only a parton-level
simulation, our phase-space cuts are conservative enough and, therefore,
the conclusions should not change significantly even on the inclusion
of hadronization effects. Initial-state radiation, which has been
neglected here, is perhaps more likely to have a discernible effect,
particularly for $N\bar{N}$ production,
but still would not change the conclusions significantly, as independent
studies have indicated. For a heavy neutrino, mass reconstruction
works very well and angular distributions can be used to determine
the quantum numbers. If the neutrino were to have a significant amount
of Majorana mass, a like-sign dilepton pair would be a very distinct
signal.

\noindent
{\bf Acknowledgement:} We acknowledge useful discussions with
G.~Giudice, T.~Medcalf, M. Mangano, R.~R\"uckl,
T.~Sj\"ostrand and S.~Shevchenko during the course of the LEP2 workshop
at CERN, which inspired this work.
Thanks are due to D. Perret-Gallix for help with GRACE.

\newpage

\section*{Appendix}
In this appendix we present the relevant formulae for
 pair-production and subsequent decay of {\em polarized} heavy fermions at
LEP2. Since the $W$-exchange diagrams can be neglected,
the production is essentially given by $s$-channel diagrams mediated
by $\gamma,Z$ exchange.
The amplitude for the process $e^-(p_1) e^+(p_2) \longrightarrow
L(p_3,s_3) \bar{L}(p_4,s_4)$ (where $L = E,N$) can then be expressed as
\be
{\cal M} = e^2 \sum_{i= \gamma,Z}
{\cal P}_i\;
   \bar{u}(p_3,s_3)\: \gamma_\mu \left(v_i^{(L)} +
               a_i^{(L)} \gamma_5 \right) \:v(p_4,s_4) \;\:
   \bar{v}(p_2)\: \gamma_\mu \left( \vei + \aei \gamma_5 \right) \:
             u(p_1) \ ,
        \label{matel}
\ee
where
\be
{\cal P}_i = (s - m_i^2 + i \Gamma_i m_i )^{-1}
         \label{propag}
\ee
and
\be
\barr{rclcrcl}
v^{(f)}_\gamma & = & Q_e &\quad& a^{(f)}_\gamma & = & 0 \\[2ex]
v^{(f)}_Z & = & \displaystyle
                \frac{  t_{3L}^{(f)} - 2Q_f \sin^2\theta_W }
                     {2{\sin \theta_W \cos \theta_W}}
   &\quad& a^{(f)}_Z & = & \displaystyle
                     \frac{-t_{3L}^f}{ 2{\sin \theta_W \cos \theta_W}}
                                 \\[2ex]
\earr
                \label{coup:g/Z}
\ee
for $f = e,L$. For convenience, we define
\be
\barr{rcl}
{\cal K}_1 & = & \displaystyle
2 \sum_{i,j} {\cal P}_i {\cal P}_j^\ast \left( \vei \vej + \aei \aej \right)
                                        \left( \vli \vlj + \ali \alj \right)
             \\[2ex]
{\cal K}_2 & = & \displaystyle
2 \sum_{i,j} {\cal P}_i {\cal P}_j^\ast \left( \vei \vej + \aei \aej \right)
                                        \left( \vli \vlj - \ali \alj \right)
             \\[2ex]
{\cal K}_3 & = & \displaystyle
2 \sum_{i,j} {\cal P}_i {\cal P}_j^\ast \left( \vei \aej + \aei \vej \right)
                                        \left( \vli \alj + \ali \vlj \right)
\ .
\earr
     \label{defn}
\ee
The production matrix element squared can then be expressed as
\be \displaystyle
 \left|{\cal M}_{\rm prod} \right|^2 = e^4 \sum_{i,j}
     \left[ {\cal A} + 4 {\cal B}_{\mu\nu} s_3^\mu s_4^\nu \right],
       \label{matelsq}
\ee
where
\be
\barr{rcl}
{\cal A} &=& \displaystyle
              {\cal K}_1
               \left[ (m_L^2 - u)^2 + (m_L^2 - t)^2 \right]
          +   {\cal K}_3  s (t - u)
          + 2 {\cal K}_2  m_L^2 s
          \\[2ex]
{\cal B}^{\mu \nu} &=& \displaystyle
                 m_L^2 {\cal K}_1
                 \left( p_1^\mu p_2^\nu + p_1^\nu p_2^\mu \right)
                + m_L^2 {\cal K}_3
                 \left( p_1^\nu p_2^\mu - p_1^\mu p_2^\nu \right)
                   \\[2ex]
          &+& \displaystyle
          \frac{ {\cal K}_2 }{2}
             \left[ (m_L^4 - u t) g^{\mu \nu}
                  + ( s - 2 m_L^2) \left( p_1^\mu p_2^\nu
                                        + p_1^\nu p_2^\mu \right)
                  + s p_3^\nu p_4^\mu
             \right.
                   \\[2ex]
          && \displaystyle  \hspace*{1em}
            \left. - p_4^\mu \left\{  (m_L^2 - u) p_1^\nu
                                    + (m_L^2 - t) p_2^\nu \right\}
                   - p_3^\nu \left\{  (m_L^2 - t) p_1^\mu
                                    + (m_L^2 - u) p_2^\mu \right\}
            \right].
\earr
          \label{defn:mat2}
\ee

The decay vertex of $L$ is of the form
$\bar{l} L W^{(*)}$, where $W^{(*)}$ is either real or virtual decaying as
$W^{(*)} \rightarrow f_1(q_1) \bar f_2(q_2)$,
$l$ is a light neutrino for $L = E$ and a light charged lepton for $L = N$.
The decay matrix element squared can then be parametrized as
\be
 \left|{\cal M}_{L \ {\rm decay} } \right|^2
       = C \left(1 + D_\mu s_3^\mu \right),
\ee
where
\be
\barr{rcl}
C & = &  \displaystyle 2 g^4 \sin^2{\theta_{lL}} \;
         \frac{ (p_3\cdot q_2)\;\; ( p_l\cdot q_1)}
        {\left[(p_3 - p_l)^2 - m_W^2\right]^2 + \Gamma_W^2 m_W^2} \\[2ex]
D_\mu & = &  \displaystyle \frac{m_L}{p_3\cdot q_2} \;\; q_{1\mu},
\earr
       \label{decay matelsq}
\ee
and $\sin{\theta_{lL}}$ is the relevant mixing angle.
Similar statements can be made for $\bar{L}$ as well.

Within the narrow-width approximation,
eqs. (\ref{matelsq} \& \ref{decay matelsq}) can then be convoluted
to give the effective matrix element squared:
\be
\barr{rcl}
\displaystyle \left|{\cal M}\right|^2_{\rm eff}
   \left(e^+ e^- \rightarrow L \bar{L}
                 \rightarrow 6 \ {\rm fermions} \right)
     =  \displaystyle
         \Gamma_L^{-2}
         \sum_{s_3, s_4}
             \left|{\cal M}_{\rm prod} \right|^2     \:
         \left|{\cal M}_{L \ {\rm decay} } \right|^2     \:
          \left|{\cal M}_{\bar{L} \ {\rm decay} } \right|^2 & &
     \\[2ex]
     =  \displaystyle
    4 e^4 \frac{C \bar{C} }{ \Gamma_L^2}
         \left[{\cal A}
               +  \frac{4}{9} {\cal B}_{\mu \nu}
                  D_\sigma \bar{D}_\lambda
                  \left( g^{\mu \sigma}
                        - \frac{ p_3^\mu p_3^\sigma}{m_L^2}  \right)
                  \left( g^{\nu \lambda}
                        - \frac{ p_4^\nu p_4^\lambda}{m_L^2}  \right)
         \right] & &
\earr
     \label{effec matelsq}
\ee
Integrating over the appropriate phase space (including the three
$\delta$-functions originating from energy-momentum conservation),
gives us the effective
cross-section. The expression in eq. (\ref{effec matelsq})
includes all spin--spin correlations.
It is easy to see that the factor
$C / \Gamma_L$, on integration, is
the same as the branching fraction of $L$
into the specific channel being considered.
Since, by definition, all decay modes for $L$ involve the factor
$\sin^2{\theta_{lL}}$, the branching fraction is independent
of this quantity.
Thus, any mixing element
at the decay vertex is irrelevant as long as it is large enough for $L$
to decay within the detector.

\end{document}